# A Motivational Driver Steering Model:

# Task Difficulty Homeostasis

# From Control Theory Perspective

H. Mozaffari, A. Nahvi

*Abstract* — **A general and psychologically plausible collision avoidance driver model can improve transportation safety significantly. Most computational driver models found in the literature have used control theory methods only, and they are not established based on psychological theories. In this paper, a unified approach is presented based on concepts taken from psychology and control theory. The "task difficulty homeostasis theory", a prominent motivational theory, is combined with the "Lyapunov stability method" in control theory to present a general and psychologically plausible model. This approach is used to model driver steering behavior for collision avoidance. The performance of this model is measured by simulation of two collision avoidance scenarios at a wide range of speeds from 20 km/h to 170 km/h. The model is validated by experiments on a driving simulator. The results demonstrate that the model follows human behavior accurately with a mean error of 7%.**

*Index Terms*— **Unified approach, Task difficulty Homeostasis Theory, Lyapunov stability method, Collision avoidance Steering behavior.**

## 1.Introduction

Cognitive science, artificial intelligence (AI), and control engineering are the main fields of science for manufacturing new generation of intelligent machines, robots, and industrial systems. Cognitive scientists (psychologists, mathematicians, and computer science researchers) try to discover human brain operations to build brain computational models. AI researchers try to investigate cognitive models or other reasonable Algorithms to make applicable intelligent systems. Control engineers try to develop mathematics of dynamic systems to grantee the stability and accuracy of the intelligent machines, robots, or other physical systems. They also consider the effect of sensor and actuator limitations in physical performance of the system. As complicated industrial intelligent systems demand all of the above abilities, today cognitive science, AI, and control engineering are getting closer so that no distinct boundary can be found between these fields of science.

Some intelligent systems, like intelligent drivers, need to represent complicate behaviors containing numerous factors, measurements, and uncertainties. Common traditional control engineering methods can not directly construct a controller to behave comprehensively, and cognitive or AI approaches should be joined to overcome the problems. Driving is an exact example of such applications especially when the intelligent system should interact with human drivers in advanced driver assistant systems (ADAS). In this study, a multidisciplinary approach is devised for intelligent driver models that can be used also in other similar intelligent systems and AI applications.

ssssThe most challenging issue to improve intelligent transportation systems is the interaction with human drivers. Intelligent systems which are not human-based probably face more complexity to understand and predict human driver's behaviors. Accordingly, a psychological plausible driver model open a new way for psychologist, AI researchers, traffic engineers, and ADAS designers to explore the human driver's behavior and interact more effectively.

Despite the fact that many researchers in the fields of traffic psychology, ergonomics, cognitive science, control theory, and traffic engineering have created numerous models to describe driver's behavior, there is no generally-accepted model yet.

In the past decades, enormous variety of driver behavior models has been introduced. According to driver model classification stated in (Winter & Happee, 2012), driver models can be classified in two major groups: unspecific models and specific models.

The first group is built on psychological point of view like motivational models (Wilde, 1994), (Fuller R. , 2005). Such models are unspecific, qualitative, and comprehensive. They do not present any mathematical formulation to be used as an input-output model. Also, there is no quantitative result to evaluate their performance (Winter & Happee, 2012). The second group is built on vehicle dynamics and control theories, which are specific, quantitative and deal with driving details. The specific models define distinct mathematical formulation to be used as an input-output model, while their paradigm is not consistent with driver psychological perspective [1]. Consequently, there is a gap between motivational and control theory models.

This paper aims to unify unspecific and specific models. A suitable mathematical formulation for motivational models is devised by control theory methods. This model enjoys the generality of motivational models and also stability and exactness of control theory models.

The risk homeostasis theory (Wilde, 1994) and the task-difficulty homeostasis theory (Fuller R. , 2005) are the most well-known motivational models. The risk homeostasis theory demonstrates that the driver controls the perceived risk and keeps it close to the target level of risk. This theory was implemented to justify why introduction of antilock brake systems did not decrease the number of accidents as had been predicted before. The task difficulty homeostasis theory considers task difficulty as the main motivation similar to the risk homeostasis theory. It is used to define how drivers choose speed in different situations.

The task-difficulty homeostasis theory is better suited to the control engineering techniques used in this paper. In this paper, task difficulty homeostasis theory is mathematically formulated by means of Lyapunov stability method (Lyapunov, 1992)of the control theory.

Task difficulty is estimated based on two major driving motivations for decision making as a) task demand, and b) driver capability. When demand is significantly less than driver capability, the task is easy. When demand equals driver capability, the task is achievable, but very difficult. If demand is more than capability, loss of vehicle control occurs. The driver tries to control the vehicle so that her/his capability exceeds task demand. This motivational model creates a platform for a general model devised in this paper.

The Lyapunov stability theory determines the stability of a dynamic system response such as a vehicle response in a collision avoidance maneuver. It defines a positive definite function and computes its time derivative. If the time derivative is negative, the positive definite function is bounded and the response is stable.

The task difficulty homeostasis theory can be formulated with the "Lyapunov stability theory" so that it can be considered as a unified multidisciplinary approach. The task demand can be used as a positive definite function. If the driver input is defined such that the time derivative of task demand becomes negative, the task demand will decrease and the loss of control will not occur.

This unified approach is applied to model driver steering behavior for collision avoidance in common traffic scenarios. Some research works have addressed driver steering models for collision avoidance using specific methods. They employ numerous methods such as fuzzy logic, (Kovacs & Koczy, 1999), (Grefe, 2005), ( Llorca , et al., 2011), and (Bauer, 2012), neural network (Chonga, Montasir, Flintschc, & Higgsd, 2013), model predictive control (MPC) (Gray , Gao, Hedrick , & Borrelli, 2013), (Kamal, JunImura , Hayakawa, Ohata , & Aiha, 2014), (Cao, Cao, Yu, & Luo, 2016), and (Erlien, Fujita , & Gerdes, 2016), and optimal control (Hayashi, Isogai, & Raksinchar, 2012), (Gordon & Gao, 2014), and (Phuc Le & Stiharu, 2013) to handle steering control. Moreover, some researchers try to consider cognitive limitations and human characteristics similar to human drivers on their models (Johns & Cole, 2012), (Johns & Cole, 2015), (Lio, et al., 2015), (Fuller, Matthew, & Liu, 2010), (Bi, Wang, Wang, & Liu, 2015), (MACADAM, 2003), (Keen & Cole, 2006), and (Odhams & Cole , 2009).

Most of the afore-mentioned models are designed to control steering wheel for specific applications. In addition, their mathematical formulation does not estimate human motivations or feelings so that they are not psychologically plausible.

Fuzzy logic driver models are usually developed for specific applications to limit the number of rules. In ( Llorca , et al., 2011), an autonomous pedestrian collision avoidance system is presented, which processes lateral displacement and speed as input signals using fuzzy logic membership functions. Although this is a valid approach, it is not based on human driver perspective and is not psychologically plausible.

Neural networks methods are also used to model driver behavior. They can fit model outputs to human driver's data. If the real traffic data are not similar to the learning data, the model may respond differently from human drivers. As general driving models need multitude of learning patterns to cover all conditions, most neural network driver models are designed for specific applications. Also, the network weight factors are not directly related to human motivations. The rule-based neural network approach in (Chonga, Montasir, Flintschc, & Higgsd, 2013) uses a four-layer rule-based neural network to find out naturalistic behavior in traffic for two situations: car following and critical event. This method predicts driver behavior appropriately in the two stated situations. However, the weight factors are not taken from human psychological perspective and the model is not general.

Model predictive and optimal controllers provide vehicles with stable maneuvers. They are based on phase plane boundaries and convex envelops and not based on driver psychology. For example, the model predictive steering controller stated in (Erlien, Fujita , & Gerdes, 2016) presents dynamic stability boundaries of a sedan to avoid obstacles. The system stated in (Gordon & Gao, 2014) finds a safe trajectory and applies optimal control to follow that trajectory. In (Phuc Le & Stiharu, 2013) a driver model is defined based on an experimental and optimal point of view. Next it is joined to vehicle dynamic for path following. Both methods are strong tools for stable vehicle control, but are not simple and psychologically plausible.

There are a few remarkable research works that use cognitive models or consider cognitive limitations. They model cognitive behavior, neuromuscular system, and visual system. Most use theoretical control models such as robust controllers (Johns & Cole, 2012), MPC controllers (Johns & Cole, 2015) and (Keen & Cole, 2006), optimal controllers (Odhams & Cole , 2009), or classical continuous transform functions (Bi, Wang, Wang, & Liu, 2015). None of these studies avoids collision and they only consider lateral control for path tracking. Besides, the mathematical formulation is not related to human motivations directly.

In this paper, a psychological theory is used to define human motivations. A control theory formulates human motivations and estimates driver's input to make a stable and psychologically plausible model.

The "task difficulty homeostasis" theory, a prominent theory in traffic psychology, is considered to define the major motivations of human driver for the collision avoidance task. Two variables of time-to-collision (TTC) and time-to-avoidance (TTA) estimate the motivational factors. A kinematic algorithm is devised to detect probable collision points and compute TTC and TTA using the relative position, velocity, and acceleration vectors. Finally, the Lyapunov theory is applied to estimate task difficulty and control steering angle with guaranteed stability.

To evaluate the model, it is simulated on a fourteen-degree-of-freedom dynamic model of a sedan in two driving scenarios. In the first scenario, the accuracy of the model is evaluated. In the second scenario, several vehicles and configurations evaluate the generality of the model. Finally, it is tested on a 90-degree turn scenario on a driving simulator to show the fidelity of the model with the behavior of the human drivers. The results confirm the expected accuracy, generality, simplicity, and validity of the model.

## 2. Modeling of collision avoidance behavior

### 2.1 Task difficulty homeostasis theory and time concepts

Based on "task difficulty homeostasis" theory, drivers always try to maintain a balance between the demands of driving task and their capability. In this model, task difficulty arises out of the comparison between task demand and capability (Fuller R. , 2005). Accordingly, drivers try to set their effort level so that the level of task difficulty is restricted in a target range. Consequently, three major concepts are defined: demand ($D$), capability ($C$), and task difficulty ($TD$).

Based on this theory, demand and capability must be compared so that the value of task difficulty can be calculated and is used as the most prominent criterion for steering control. First, $D$ and $C$ are calculated based on position, velocity, and acceleration vectors of the intelligent car and obstacles. Then $TD$ is extracted from the $D$ and $C$ values to determine the level of difficulty of the task and adjust the steering angle accordingly.

The most challenging Issue to formulate task difficulty homeostasis theory is finding two measureable quantities which can estimate the driving demand and the driver capability. These quantities are not defined mathematically in task-difficulty homeostasis theory.

The Author of the task-difficulty theory relates task demand to workload. He verifies his model in drivers' speed control for different situations. He stated that drivers can provide enough time for their needed workload with decreasing the speed (Fuller R. , 2005). In contrast, when the needed workload is low, they increase their speed to decrease the available time. As a result, high workload means short available time and low work load means the long available time. So workload and available time have a strong inverse relationship. As simple modeling of workload cannot be made computationally, available time would play a significant role to estimate the task demand in collision avoidance scenarios.

Other studies like (Summala, 1980) and (Hollnagel, 2002) also suggest using available time in psychological driver models. Safety margin model (Summala, 1980) defines safe zones based on available time. In this way the actions which provide enough available times are safe and the drivers can select the safe action which is more comfortable. The conceptual control model (COCOM) (Hollnagel, 2002) also modeled the drivers' control behavior in a time-cycle. The available time is compared with the time to evaluate an event, time to select an action, and time to perform the action. Drivers behave so that the available time would be greater than the sum of them.

Hence, the available time would be a suitable quantity to model the task demand. Time to collision *TTC* imply how much time is available for driver to act in collision avoidance scenarios. Thus *TTC* can be used logically to model the demand in the collision avoidance behavior model.

*TTC* have also another logical benefit. This quantity can define which obstacle has a higher priority for avoidance. The obstacle colliding sooner than the others should be assigned a higher priority for avoidance.

A controversial issue still remains. Does the time to collision have a high correlation to human drivers control actions? In (Van Der Horst, 2007), the author tries to measure driving risk based on longitudinal time to collision and lateral time to lane crossing. He finds high correlation between longitudinal time to collision with drivers' brake and accelerator pedal rates. Also in lateral control for lane keeping, he shows time to lane crossing has a high correlation to drivers' steering angle. In this way using longitudinal time to collision and lateral time to lane crossing seems efficient in driver behavior modeling. Both longitudinal time to collision and lateral time to lane crossing can be replaced by combined longitudinal and lateral time to collision *TTC*. Although *TTC* has not been formulated for combined lateral and longitudinal situations yet, it can be computed for all positions, velocities, and accelerations vectors by an algorithm devised in this paper. In this way, *TTC* can evaluate the task demand comprehensively.

All in all, *TTC* can inversely determine the demand *D* for avoiding collision. Multitude mathematical functions can formulate such an inverse relationship. There are some conditions that selected function should satisfy.

The demand should be infinity when TTC is zero. Lyapunov theory certifies that the demand remains bounded, and when demand is bounded TTC is not zero and collision would not occur. This condition certifies bounded demands lead no colliding. Similarly, when TTC goes infinity, demand should get close to zero. No extra action is needed when TTC is very long. Also, simple functions are preferred. Several functions are tested and the following simple formula has better results:

$$D = 1/TTC \qquad\qquad (1)$$

*TTC* between two collision points can be easily estimated by the following first-order estimation:

$$TTC = \begin{cases} -\dfrac{S}{\dot{S}} & \dot{S} < 0 \\ \infty & \dot{S} \geq 0 \end{cases} \qquad\qquad (2)$$

where $S$ is the distance between two colliding points. In order to apply S and its temporal rate $\dot{S}$ in equation (2), the colliding points should be defined. In this paper, an algorithm is presented to detect probable collision points and estimate the demand for moving obstacles as described in section 2.2.

Driver Capability also should be estimated by a measurable quantity. It is preferred capability is estimated by a time-related quantity to simplify the comparison between demand and capability.

In task-difficulty homeostasis theory capability is a feed-forward and self-appraised quantity. Drivers have a predefined appraise of their ability (Fuller R. , 2005). In this paper, capability is defined based on feed-forward and feed-back monitoring. We found that drivers evaluate a predefined capability in the beginning of the event, and then they appraise their capability continuously by feed-back monitoring. After they turn the steering wheel angle, they estimate how much their effort is effective. Thus, a feedback signal similar to time to collision can model driver-vehicle capability.

A capable driver can apply large lateral and longitudinal accelerations to prevent the vehicle from collision. A capable driver can change the steering angle in a short period of time or press on the brake pedal quickly. The relative acceleration $\ddot{S}$ depends on the effort of the driver. The driver decides on the appropriate amount of acceleration based on the estimated $\dot{S}$. If the time the driver needs to make $\dot{S}$ zero is less than TTC, the driver is capable. If the time the driver needs to make $\dot{S}$ zero is more than TTC, the driver is incapable. Thus the time needed to make $\dot{S}$ zero can be used to estimate driver's capability.

 Driver capability leads to decline or improvement of the situation. If the driver is not capable enough, task demand increases; if the driver is capable enough, task demand decreases. We conclude that the capability is related to the time derivative of the task demand. Accordingly, a proper time-related quantity may be extracted from the time derivative of the task demand:

$$\frac{dD}{dt} = \frac{d\left(\frac{-\dot{S}}{S}\right)}{dt} = -\frac{\ddot{S}S - \dot{S}^2}{S^2} = \frac{\dot{S}^2}{S^2} - \left(\frac{\ddot{S}}{-\dot{S}}\right)\left(\frac{-\dot{S}}{S}\right) \qquad (3)$$

Thus,

$$\frac{dD}{dt} = D^2 - \left(-\frac{\ddot{S}}{\dot{S}}\right)D = D\left(D - \left(-\frac{\ddot{S}}{\dot{S}}\right)\right) \qquad (4)$$

Equation (4) shows that the temporal rate of task demand is related to $D$ and also $(-\ddot{S}/\dot{S})$. The term $(-\ddot{S}/\dot{S})$ determines whether the task demand increases or decreases. So this term is a reasonable candidate for capability. The inverse of this term, that is $(-\dot{S}/\ddot{S})$, is a first order estimation of the time needed for setting the relative velocity $\dot{S}$ to zero. If $\dot{S}$ becomes zero, there will be no collision.

This temporal quantity can predict how long it will take to avoid collision. It can be termed as "Time-To-Avoidance" ($TTA$):

$$TTA = \begin{cases} -\dot{S}/\ddot{S} & \ddot{S} > 0 \\ \infty & \ddot{S} \leq 0 \end{cases} \qquad (5)$$

The longer the $TTA$, the less capable the driver will be. Similar to demand formula (equation (1)), the capability is computed by:

$$C = 1/TTA \qquad (6)$$

It is noticeable that D and C has the same unit ($s^{-1}$). So the subtract operator can compare these quantities to estimate task-difficulty.

Capability is meaningful when demand exists. If no obstacle is present nearby or if it is moving away from the intelligent vehicle, demand will be zero and capability is set to zero.

### 2.2 D and C estimation by kinematic analysis

To compute $S$, it is necessary to detect potential collision points. The vehicles and the moving obstacles are bounded by rectangles. The upcoming collision points of these moving rectangles should be located.

The vertex of one rectangle collides with the edge of another when collision occurs. To detect the collision points, a line is drawn parallel to the direction of the relative velocity of the vehicles from each vertex. The intersection point of each line with the edges of another rectangle indicates a candidate colliding point. For example, as shown in **Fig. 1**, a line is drawn from the vertex $C$ of the obstacle parallel to the relative velocity $V_{rel}$, which crosses the edge of the vehicle at the point $P$. The most likely colliding point lies on a line with the shortest TTC such as line segment $PC$ in Fig. 1. The velocity directions of the intelligent vehicle $V_v$ and the obstacle $V_o$ are assumed to be equal to the tangent of steering angle. This is an accurate assumption if the tires do not have large lateral slips. If the obstacle steering angle is not accessible, the obstacle velocity vector can be used instead.

To find out the colliding points between the intelligent car and the road edge, the algorithm is similar to that used for the intersection points of the rectangles. Four lines parallel to the vehicle steering direction are drawn from the vertices of the intelligent vehicle. The intersections of these lines with the road edge are candidate collision points. The collision point having the smallest TTC is the most possible collision point like point $C_r$ Fig. 1.

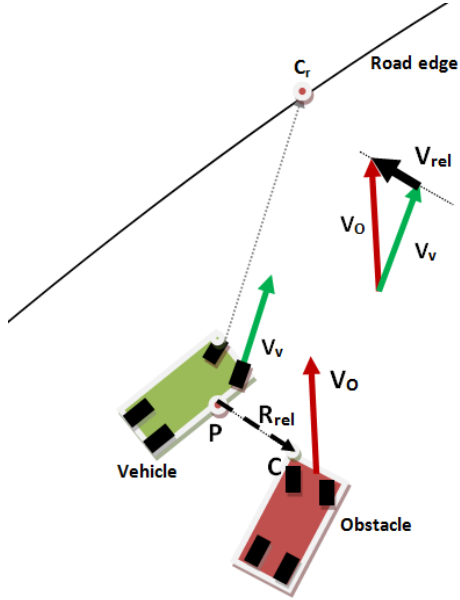

Fig. 1 Detection of colliding points in a typical traffic configuration. The dotted line shows the direction of relative velocity

As shown in **Fig. 2**, $R_{rel}$ is the relative position vector between the colliding points. The distance between the colliding points $S$ is equal to the magnitude of $R_{rel}$ and can be calculated as follows:

$$S = \sqrt{\vec{R}_{rel} . \vec{R}_{rel}} \tag{7}$$

The time derivative of $S$ is obtained as follows:

$$\dot{S} = \frac{\dot{\vec{R}}_{rel} . \vec{R}_{rel}}{\sqrt{\vec{R}_{rel} . \vec{R}_{rel}}} \tag{8}$$

where $\dot{\vec{R}}_{rel}$ is the relative velocity between colliding points. Substituting equations (7) and (8) into equation (2) and (1), task demand can be estimated as:

$$D = \begin{cases} -\dfrac{\dot{\vec{R}}_{rel} . \vec{R}_{rel}}{\vec{R}_{rel} . \vec{R}_{rel}} & \dot{\vec{R}}_{rel} . \vec{R}_{rel} < 0 \\ 0 & \dot{\vec{R}}_{rel} . \vec{R}_{rel} \geq 0 \end{cases} \tag{9}$$

$\ddot{S}$ is obtained by the time derivative of equation (8) as follows:

$$\ddot{S} = \frac{\left( \dot{\vec{R}}_{rel} . \dot{\vec{R}}_{rel} + \ddot{\vec{R}}_{rel} . \vec{R}_{rel} \right)}{\sqrt{\vec{R}_{rel} . \vec{R}_{rel}}} - \left( \frac{\dot{\vec{R}}_{rel} . \vec{R}_{rel}}{\sqrt{\vec{R}_{rel} . \vec{R}_{rel}}} \right) \left( \frac{\dot{\vec{R}}_{rel} . \vec{R}_{rel}}{\vec{R}_{rel} . \vec{R}_{rel}} \right) \tag{10}$$

Substituting equations (8) and (10) in equations (5) and (6) yields:

$$C = \begin{cases} \max\left( 0, \left[ -\dfrac{\dot{\vec{R}}_{rel} . \dot{\vec{R}}_{rel}}{\dot{\vec{R}}_{rel} . \vec{R}_{rel}} - \dfrac{\ddot{\vec{R}}_{rel} . \vec{R}_{rel}}{\dot{\vec{R}}_{rel} . \vec{R}_{rel}} - D \right] \right) & D > 0 \\ 0 & D \leq 0 \end{cases} \tag{11}$$

As shown in **Fig. 2**, the relative position vector should be obtained as follows:

$$\vec{R}_{rel} = \left( \vec{R}_o + \vec{R}_{oc} \right) - \left( \vec{R}_v + \vec{R}_{vp} \right)$$

$$\vec{R}_{rel} = \left[ \begin{bmatrix} X_O \\ Y_O \end{bmatrix} + \begin{bmatrix} \cos\psi_O & -\sin\psi_O \\ \sin\psi_O & \cos\psi_O \end{bmatrix} \begin{bmatrix} a_{oc} \\ b_{oc} \end{bmatrix} \right]$$
$$- \left[ \begin{bmatrix} X_v \\ Y_v \end{bmatrix} + \begin{bmatrix} \cos\psi_v & -\sin\psi_v \\ \sin\psi_v & \cos\psi_v \end{bmatrix} \begin{bmatrix} a_{vp} \\ b_{vp} \end{bmatrix} \right] \qquad (12)$$

where $X_o$ and $Y_o$ are position coordinates of the obstacle center in the reference coordinate system (X,Y), $\psi_O$ is the yaw angle of the obstacle, and $a_{oc}$ and $b_{oc}$ are position coordinates of the obstacle collision point in the obstacle coordinate system as shown **Fig. 3**. $X_v$ and $Y_v$ are position coordinates of the vehicle center of mass in the reference coordinate system, $\psi_v$ is the yaw angle of the vehicle, $a_{vp}$ and $b_{vp}$ are position coordinates of the vehicle collision point in the vehicle coordinate system as shown in **Fig. 4**.

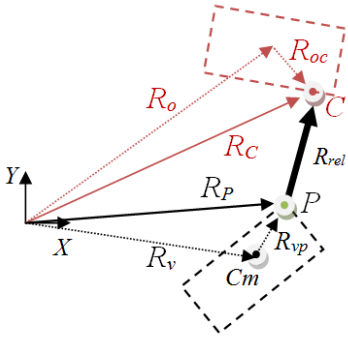

Fig. 2. Geometric illustration of position vector for the potential collision points of the vehicle and the obstacle

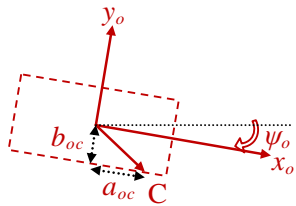

Fig. 3. Obstacle coordinate system and the location of collision point

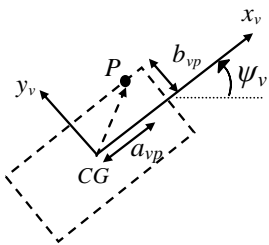

Fig. 4. Vehicle coordinate system and the location of collision point

To calculate the relative velocity between collision points, the time derivative of equation (12) is obtained as follows:

$$\dot{R}_{rel} = \left[\begin{bmatrix} \dot{X}_O \\ \dot{Y}_O \end{bmatrix} + \begin{bmatrix} -\sin\psi_O & -\cos\psi_O \\ \cos\psi_O & -\sin\psi_O \end{bmatrix}\begin{bmatrix} a_{oc} \\ b_{oc} \end{bmatrix}\dot{\psi}_O\right]$$
$$- \left[\begin{bmatrix} \dot{X}_v \\ \dot{Y}_v \end{bmatrix} + \begin{bmatrix} -\sin\psi_v & -\cos\psi_v \\ \cos\psi_v & -\sin\psi_v \end{bmatrix}\begin{bmatrix} a_{vp} \\ b_{vp} \end{bmatrix}\dot{\psi}_v\right]$$

(13)

To find out relative acceleration vector $\ddot{R}_{rel}$, the time derivative of equation (13) yields:

$$\ddot{R}_{rel} = \left[\begin{array}{l}\begin{bmatrix} \ddot{X}_O \\ \ddot{Y}_O \end{bmatrix} + \begin{bmatrix} -\cos\psi_O & \sin\psi_O \\ -\sin\psi_O & -\cos\psi_O \end{bmatrix}\begin{bmatrix} a_{oc} \\ b_{oc} \end{bmatrix}\dot{\psi}_O{}^2 \\ + \begin{bmatrix} -\sin\psi_O & -\cos\psi_O \\ \cos\psi_O & -\sin\psi_O \end{bmatrix}\begin{bmatrix} a_{oc} \\ b_{oc} \end{bmatrix}\ddot{\psi}_O\end{array}\right]$$
$$- \left[\begin{array}{l}\begin{bmatrix} \ddot{X}_v \\ \ddot{Y}_v \end{bmatrix} + \begin{bmatrix} -\cos\psi_v & \sin\psi_v \\ -\sin\psi_v & -\cos\psi_v \end{bmatrix}\begin{bmatrix} a_{vp} \\ b_{vp} \end{bmatrix}\dot{\psi}_v{}^2 \\ + \begin{bmatrix} -\sin\psi_v & -\cos\psi_v \\ \cos\psi_v & -\sin\psi_v \end{bmatrix}\begin{bmatrix} a_{vp} \\ b_{vp} \end{bmatrix}\ddot{\psi}_v\end{array}\right]$$

(14)

In summary, the potential collision points are found by the method illustrated in Fig. 1 for all corners of the obstacle and the vehicle. The relative position, relative velocity, and relative acceleration vectors are calculated by equations (12), (13), and (14)respectively. Finally, task demand is estimated by equation (9) and driver capability is estimated by equation (11). The point with maximum task demand $D$ is the most likely collision point.

### 2.3 Numerical estimation of task difficulty based on Lyapunov stability theory

To determine the task difficulty, Lyapunov stability method is applied. In this method, a positive definite function is considered. An input is then defined to make the time derivative of that function negative definite (Lyapunov, 1992). The positive definite function of such a system will be bounded.

The aim of this study is to define the steering angle such that no collision occurs. Half of task demand squared is chosen as a Lyapunov positive definite function:

$$V = \frac{1}{2}D^2 = \frac{1}{2}\left(\frac{\dot{S}}{S}\right)^2$$

(15)

If the Lyapunov function remains bounded, the collision distance $S$ cannot become zero and the accident will not occur.

The time derivative of the Lyapunov function is calculated using equation (9):

$$\frac{dV}{dt} = D^2\left(D + \frac{\ddot{\vec{R}}_{rel}.\vec{R}_{rel}}{\dot{\vec{R}}_{rel}.\vec{R}_{rel}} + \frac{\dot{\vec{R}}_{rel}.\dot{\vec{R}}_{rel}}{\dot{\vec{R}}_{rel}.\vec{R}_{rel}} + D\right) < 0$$

(16)

Based on **Fig. 2**, $\ddot{\vec{R}}_{rel}$ is defined as:

$$\ddot{\vec{R}}_{rel} = \ddot{\vec{R}}_c - \ddot{\vec{R}}_p \qquad (17)$$

Replacing above equation in equation (16) yields:

$$\left( D + \frac{\left( \ddot{\vec{R}}_c - \ddot{\vec{R}}_p \right).\vec{R}_{rel}}{\dot{\vec{R}}_{rel}.\vec{R}_{rel}} + \frac{\dot{\vec{R}}_{rel}.\dot{\vec{R}}_{rel}}{\dot{\vec{R}}_{rel}.\vec{R}_{rel}} + D \right) < 0 \qquad (18)$$

It is necessary to include the driver input (steering angle) in above equation. First the acceleration vector of point P shown in **Fig. 4** is defined as:

$$\ddot{\vec{R}}_p = \ddot{\vec{R}}_v + \begin{bmatrix} -a_{vp} \\ -b_{vp} \end{bmatrix} \dot{\psi}_v^{\,2} + \begin{bmatrix} -b_{vp} \\ a_{vp} \end{bmatrix} \ddot{\psi}_v \qquad (19)$$

where $\ddot{\vec{R}}_v$ is the vehicle center of mass acceleration vector expressed in the vehicle coordinate system. The nonlinear equations of a planar vehicle dynamics is written as follows:

$$\ddot{\vec{R}}_v = \begin{bmatrix} \ddot{x} - \dot{y}\dot{\psi} \\ \ddot{y} + \dot{x}\dot{\psi} \end{bmatrix} = \vec{f}(Z,\delta) \qquad (20)$$

$$Z = \begin{bmatrix} \psi_v, \dot{\psi}_v, \dot{x}_v, \dot{y}_v \end{bmatrix}$$

$$\ddot{\psi}_v = g(Z,\delta) \qquad (21)$$

where $Z$ is the state vector, $\dot{x}_v$ and $\dot{y}_v$ are the longitudinal and lateral speed of the vehicle expressed in the vehicle coordinate system, and $\delta$ is the steering angle.

Replacing equation (20) and (21) in equation (19) yields:

$$\ddot{\vec{R}}_p = \vec{f}(Z,\delta) + \begin{bmatrix} -a_{vp} \\ -b_{vp} \end{bmatrix} \dot{\psi}_v^{\,2} + \begin{bmatrix} -b_{vp} \\ a_{vp} \end{bmatrix} g(Z,\delta) \qquad (22)$$

Defining new function called $h$ yields:

$$\vec{h}(Z,\delta) = \vec{f}(Z,\delta) + \begin{bmatrix} -a_{vp} \\ -b_{vp} \end{bmatrix} \dot{\psi}_v^{\,2} + \begin{bmatrix} -b_{vp} \\ a_{vp} \end{bmatrix} g(Z,\delta) \qquad (23)$$

$$\ddot{\vec{R}}_p = \vec{h}(Z,\delta)$$

In driving process, drivers do not know the exact position of the steering wheel and they only may control the steering wheel speed (Tan & Huang, 2011). In discrete form, we can conclude that they control steering angle change $\Delta\delta$. In conclusion, the vehicle acceleration should be found for $\delta + \Delta\delta$. To develop the Lyapunov method for discrete form, a Taylor's estimation is applied as below:

$$\ddot{\vec{R}}_p(Z^k,\delta^{k+1}) = \ddot{\vec{R}}_p(Z^k,\delta^k) + \frac{\partial \vec{h}}{\partial \delta}\bigg|_{\delta^k} \Delta\delta + O^2 + \dots \qquad (24)$$

If the order of the magnitude of second order and higher order terms can be neglected in comparison with first order one, it can be estimated that:

$$\ddot{\vec{R}}_p(Z, \delta + \Delta\delta) \approx \ddot{\vec{R}}_p(Z, \delta) + \frac{\partial \vec{h}}{\partial \delta}\bigg|_{\delta} \Delta\delta \qquad (25)$$

Replacing equation (25) in equation (18) yields:

$$\left( D + \frac{\left( \ddot{\vec{R}}_c - \ddot{\vec{R}}_p\big|_{\delta} - \frac{\partial \vec{h}}{\partial \delta} \Delta\delta \right).\vec{R}_{rel}}{\dot{\vec{R}}_{rel}.\vec{R}_{rel}} + \frac{\dot{\vec{R}}_{rel}.\ddot{\vec{R}}_{rel}}{\dot{\vec{R}}_{rel}.\vec{R}_{rel}} + D \right) < 0 \qquad (26)$$

The final condition of stability can be expressed as:

$$\left( D - \left( -\frac{\ddot{\vec{R}}_{rel}\big|_{\delta}.\vec{R}_{rel}}{\dot{\vec{R}}_{rel}.\vec{R}_{rel}} - \frac{\dot{\vec{R}}_{rel}.\ddot{\vec{R}}_{rel}}{\dot{\vec{R}}_{rel}.\vec{R}_{rel}} - D \right) - \frac{\frac{\partial \vec{h}}{\partial \delta}.\vec{R}_{rel}}{\dot{\vec{R}}_{rel}.\vec{R}_{rel}} \Delta\delta \right) < 0 \qquad (27)$$

Based on equation (11), equation (27) can be rewritten as:

$$\left( D - C \right) - \frac{\frac{\partial \vec{h}}{\partial \delta}.\vec{R}_{rel}}{\dot{\vec{R}}_{rel}.\vec{R}_{rel}} \Delta\delta \prec 0 \qquad (28)$$

The minimum required $\Delta\delta$ is:

$$\Delta\delta = \begin{cases} \frac{\dot{\vec{R}}_{rel}.\vec{R}_{rel}}{\frac{\partial \vec{h}}{\partial \delta}.\vec{R}_{rel}} \left( D - C \right) & D - C > 0 \\ 0 & D - C = 0 \end{cases} \qquad (29)$$

Based on equation (29), task difficulty can be defined as below:

$$TD = \begin{cases} D - C & D - C > 0 \\ 0 & D - C < 0 \end{cases} \qquad (30)$$

Steering control gain $K_s$ is defined as:

$$K_s = \frac{\dot{\vec{R}}_{rel}.\vec{R}_{rel}}{\frac{\partial \vec{h}}{\partial \delta}.\vec{R}_{rel}} \qquad (31)$$

The final $\Delta\delta$ can be defined as:

$$\Delta\delta = K_s TD \qquad (32)$$

To estimate $\partial h / \partial \delta \cdot \vec{R}_{rel}$ in equation (31), the vehicle dynamics should be considered. According to equation (23), the inner product of equation (29) can be rewritten in scalar form as below:

$$\frac{\partial \vec{h}}{\partial \delta} \cdot \vec{R}_{rel} = \left( \frac{\partial f_x}{\partial \delta} - b_{vp} \frac{\partial g}{\partial \delta} \right) x_{rel} + \left( \frac{\partial f_y}{\partial \delta} + a_{vp} \frac{\partial g}{\partial \delta} \right) y_{rel} \quad (33)$$

where $x_{rel}$ and $y_{rel}$ are longitudinal and lateral components of $R_{rel}$ in vehicle coordinate system.

To simplify mathematical formulation, the bicycle model is preferred. In **Fig. 5** the free diagram of car body is represented. Applying second newton's law, the acceleration components can be obtained as below:

$$f_x = \frac{1}{m} \left( F_{f_t} \cos(\delta) - F_{f_n} \sin(\delta) + F_{r_t} - \tfrac{1}{2} c_d \dot{x}^2 \right) \quad (34)$$

$$f_y = \frac{1}{m} \left( F_{f_t} \sin(\delta) + F_{fn} \cos(\delta) + F_{rn} \right) \quad (35)$$

Also by using moments of forces, $g(Z,\delta)$ is defined as:

$$g = \frac{1}{J} \left( F_{f_t} l_f \sin(\delta) + F_{f_n} l_f \cos(\delta) - F_{r_n} l_r \right) \quad (36)$$

where $l_f$ and $l_r$ are the distance between the front and rear wheels and vehicle center of mass. Tire forces are related to the tangential and normal slip of front and rear tires and the steering angle. The slips of tires are related to vehicle velocity and vehicle yaw rate. In low slip angle a linear relationship can be considered as bellow:

$$F_{rt} = C_{rt} \beta_{rt} \quad (37)$$

$$F_{ft} = C_{ft} \beta_{ft} \quad (38)$$

$$F_{rn} = C_{rn} \beta_{rn} \quad (39)$$

$$F_{fn} = C_{fn} \left( \beta_{fn} - \delta \right) \quad (40)$$

where $C_{rt}$ and $C_{ft}$ are rear and front tangential tire stiffness respectively. $C_{rn}$ and $C_{fn}$ are rear and front normal tire stiffness respectively. $\beta_{rt}$ and $\beta_{ft}$ are rear and front tangential slip percent. $\beta_{rn}$ and $\beta_{fn}$ are rear and front normal slip angles. The Moments of tires are not considerable.

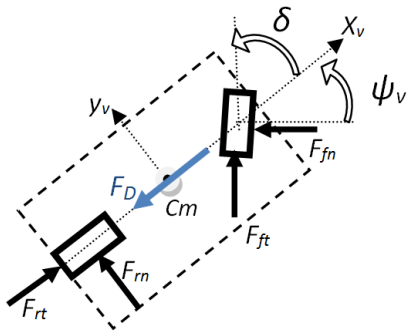

Fig. 5. Free Diagram of planar vehicle dynamics

The tangential force may be neglected when brake or accelerator pedal are not pushed. By applying equations (37) - (40) in equations (34) - (36) and making derivative, yields:

$$\frac{\partial f_x}{\partial \delta} = \frac{C_{fn}}{m}\left(\sin(\delta) - (\beta_{fn} - \delta)\cos(\delta)\right) \qquad (41)$$

$$\frac{\partial f_y}{\partial \delta} = -\frac{C_{fn}}{m}\left(\cos(\delta) + (\beta_{fn} - \delta)\sin(\delta)\right) \qquad (42)$$

$$\frac{\partial g}{\partial \delta} = -\frac{C_{fn} l_f}{J}\left(\cos(\delta) + (\beta_{fn} - \delta)\sin(\delta)\right) \qquad (43)$$

Lateral front slip angle $\beta_{fn}$ is defined as follows:

$$\beta_{fn} = \arctan\left(\frac{\dot{y}_v + l_f \dot{\psi}}{\dot{x}_v}\right) \qquad (44)$$

If front wheel lateral slip angle is not accessible, it can be neglected with no considerable accuracy loss.

All formulation to compute driver input is presented. First task difficulty is defined from equation (30). Then $\frac{\partial \vec{h}}{\partial \delta} \cdot \ddot{\vec{R}}_{rel}$ can be calculated by equations (41)-(43) and equation (33). The $K_s$ is computed by equation (31). Finally steering angle change is defined based on equation (32).

### 2.4 Collision avoidance algorithm

When a few obstacles are present, only the most demanding left obstacle and the most demanding right obstacle are considered. The desired steering angle change is obtained by the difference between this left and right requirement. For a right obstacle $i$, the change in the steering angle $\Delta \delta_i$ is counterclockwise and positive. Among all right obstacles, the most demanding obstacle requires the maximum positive steering angle change, *i.e.* $\max(0, \Delta \delta_i)$. Similarly, for a left obstacle $i$, the change in the steering angle $\Delta \delta_i$ is clockwise and negative. Among all left obstacles, the most demanding obstacle requires the minimum negative steering angle change, *i.e.* $\min(0, \Delta \delta_i)$. In the presence of $N$ obstacles, the steering angle considers both sides and is obtained by:

$$\Delta \delta = \max(0, \Delta \delta_i) + \min(0, \Delta \delta_i) \qquad i = 1 : N \qquad (45)$$

The collision avoidance algorithm consists of the following steps:

1. Find collision points for all obstacles according to presented method in Fig. 1. The $R_{vp}$ and $R_{oc}$ must be computed.

2. Define $\vec{R}_{rel}, \dot{\vec{R}}_{rel}$, and $\ddot{\vec{R}}_{rel}$ for all obstacles by simulation data in traffic simulations or sensory data of the intelligent vehicle using equations (12),(13) and (14).

3. Calculate $D_i$, $C_i$, and $TD_i$ by equations (9), (11) and (30) for all obstacles.

4. Calculate $\frac{\partial \vec{h}}{\partial \delta} \cdot \dot{\vec{R}}_{rel}$ by equations (41), (43) and (33).

5. Calculate $K_s$ by equation (31)

6. Calculate $\Delta \delta_i$ by equation (32) for all obstacles.

7. Calculate $\Delta \delta$ by equation (45)

8. Define new steering angle by the following equation

$$\delta^{k+1} = \delta^k + \Delta \delta \qquad (46)$$

## 3. Simulation results

In this section, the accuracy and generality of the driver model is evaluated by simulation of two driving scenarios. A nonlinear dynamic model of the vehicle is used to simulate the motion of the vehicle.

In the first scenario, the positioning accuracy of the driver model is evaluated. A moving obstacles force the driver model to avoid collision by moving the vehicle into a tight space.

In the second scenario, the performance of the model in the presence of several obstacles is evaluated. Four obstacles with different sizes move in both lateral and longitudinal directions and force the driver model to avoid collision.

### 3.1 Positioning the vehicle in a moving tight space

In this scenario, a moving obstacle is driving close to the vehicle. Then it turns toward the vehicle (**Fig. 6**) so that only a tight space remains between the vehicle and the obstacles. If the driver model cannot adjust the vehicle direction and the lateral position simultaneously, collision would be unavoidable. This scenario can estimate the accuracy of the model.

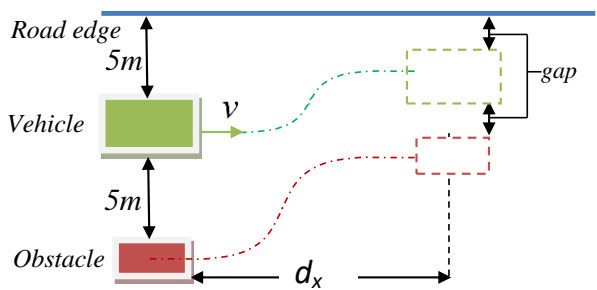

Fig. 6. A moving obstacles force the driver model to avoid collision by driving the vehicle in a tight gap.

The left obstacle is road edge. The right obstacle is a *3.6m×1.6m* sedan. The vehicle is also a *4.4m×1.7m* typical sedan. The obstacles are chosen different to make the scenario more general.

At first, the vehicle and the moving obstacle travel parallel to each other with the lateral distance of *5 m* with the same speed of *$V_x$*. Then, the right obstacle turns towards the vehicle according to a sigmoid function. This function is defined such that the lateral movement is completed after traveling a horizontal distance $d_x$. The vehicle should be able to position itself within a small gap between the obstacles. The *gap*, the speed $V_x$, and the horizontal distance $d_x$ are the parameters, which determine the difficulty of the scenario.

The simulation is run for *gap=20, 40, 60, 90 cm*, $d_x$*=20, 30, 45, 60 m*, and different speeds. The maximum speed at which the model is capable of avoiding collision is found as shown in Table 1.

Table 1
Maximum speed (KM/h) for different conditions (Ts=1/24 s)

| $d_x$<br>gap | 20 m<br>very steep lateral motion | 30 m<br>steep lateral motion | 40 m<br>moderate lateral motion | 50 m<br>mild lateral motion | 60 m<br>very mild lateral motion |
|---|---|---|---|---|---|
| 20 cm<br>extra tight gap | 45 | 75 | 95 | 110 | 125 |
| 40 cm<br>very tight gap | 50 | 80 | 100 | 125 | 145 |
| 60 cm<br>tight gap | 60 | 85 | 105 | 130 | 155 |
| 90 cm<br>moderate gap | 70 | 90 | 115 | 140 | 170 |

The results show that the driver model avoids obstacles for a vast range of speeds. As the horizontal distance increases, the lateral movements are milder and easier. Thus the maximum speed at which the driver model is capable of avoiding collision, increases. As the gap increases, the vehicle has more space to move, and higher speeds are achievable. Thus, speeds at the bottom and right cells of Table 1 are higher than those at the top and left cells. For example, the driver model avoids collision at the tight gap of *40 cm* and short horizontal distance of *30 m* for the maximum speed of *80 km/h*, which is reasonable for urban streets. But, for the moderate gap of *90 cm* and moderate horizontal distance of *40 m*, the maximum speed of *115 km/h* is reachable with no collision, which is reasonable for highways.

The results for *gap=20 cm, $d_x$=50 m,* and $V_x$*=110 km/h* are shown in Fig. **7**-Fig. **14**.

The car direction and lateral position should be controlled by steering wheel simultaneously so that the target vehicle can position itself between the two obstacles in a very tight and moving space. The driver model can locate the vehicle in a tight gap of *20 cm,* while the length of the vehicle is *4.4 m* (Fig. **7**).

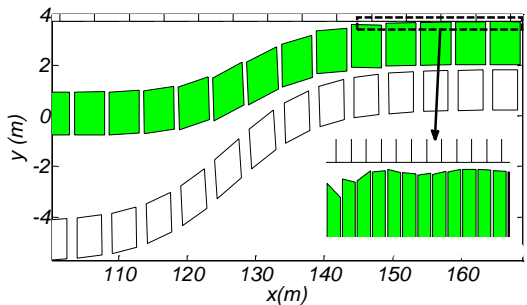

Fig. **7**. Trajectory of vehicles in gap=20cm, d=50(m), and V= 110 km/h. The vehicle is shown by a solid box and the obstacle is shown by a hollow box. The gap between road edge and the vehicle is magnified to show no collision occurs.

The maximum task demand can show how many times the controller can set the steering angle change before collision. For example, the maximum task demand is $6s^{-1}$ as shown in Fig. **8**. The minimum *TTC* is *1/6 s*. As the sample time is *1/24 s*, the controller can set the steering angle four times before collision occurs. Decreasing the sample time increases the driver model accuracy.

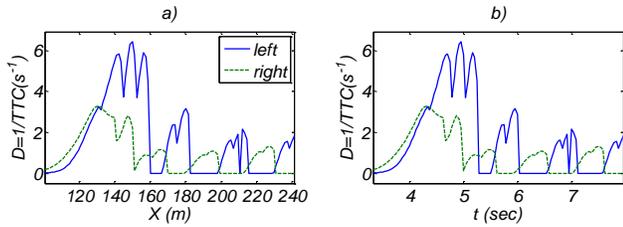

Fig. **8**. The level of task demand for both sides in gap=20cm, d=50 (m), and V= 110 km/h. a) Task demand-horizontal location. b) Task-demand-time.

The driver model capability is plotted for the left and right obstacle in Fig. **9**. The maximum capability is about *2 s^{-1}*, which is less than maximum task demand. As the two obstacles are located at both sides of the vehicle, the driver model cannot change the steering angle as long as capability exceeds task demand. Task difficulty is shown in Fig. **10** for the left and right obstacle.

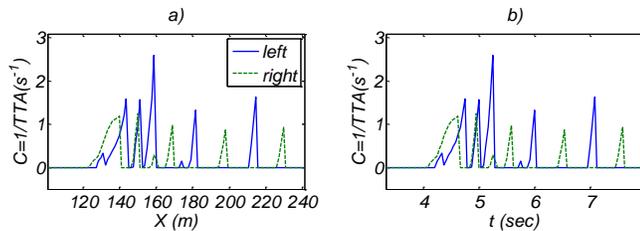

Fig. **9**. Capability for both sides in gap=20cm, d=50 (m), and V= 110 km/h. a) Capability-horizontal location. b) Capability-time.

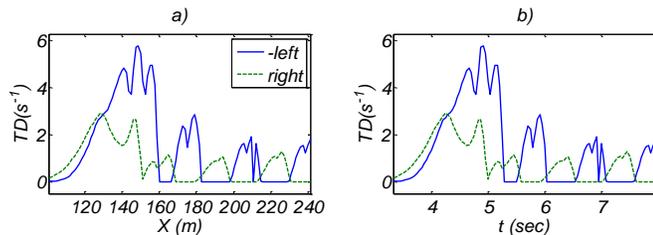

Fig. **10**. The task difficulty for both sides in gap=20cm, d=50m, and V= 110 km/h. a) Task difficulty-Horizontal location. b) Task difficulty-time.

The driver steering wheel angle, which is 16 times larger than the tire steering angle, is plotted in Fig. **11**, Fig. **12**, and Fig. **13**. The steering wheel change for left and right obstacle is plotted in Fig. **11**. The semi-symmetric shape indicates that the driver model can balance the steering wheel change for left and right obstacles.

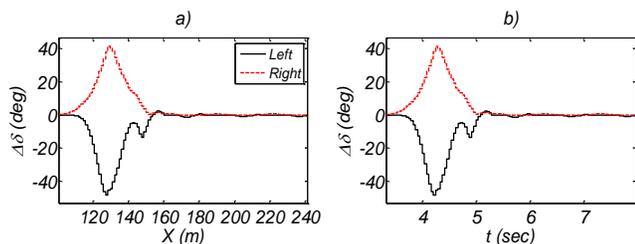

Fig. **11**. The steering wheel angle change for left and right obstacles in gap=20cm, d=50 m, and V= 110 km/h .a) Steering wheel angle change-horizontal location. b) Steering wheel angle change-time.

According to equation (45), the driver model steering wheel change is plotted in Fig. **12**. To verify that the driver model steering wheel change is reasonable, the Ackerman steering wheel change is also plotted in that figure. The Ackerman steering wheel angle can be estimated by locating the vehicle in the middle of the two obstacles. Note that the Ackerman estimation can avoid collision only for very small slip angles.

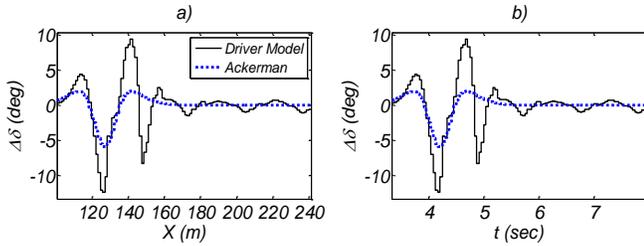

Fig. **12**. The steering wheel angle change in gap=20cm, d=50 m, and V= 110 km/h. a) Steering wheel angle change -horizontal location. b) Steering wheel angle change -time.

Similarly, the driver model steering wheel angle and the Ackerman steering wheel angle are compared in Fig. **13**. There is a difference between Ackerman and driver model steering wheel angle.

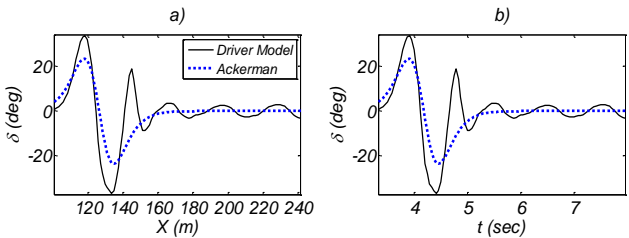

Fig. **13**. The exerted steering wheel angle change in gap=20cm, d=50 m, and V= 110 km/h. a) Steering wheel angle-horizontal location. b) Steering wheel angle–time. There is difference between Ackerman and driver model steering wheel angle.

This difference is reasonable because the lateral slip angle is considerable as shown in Fig. **14** and exceeds *2 deg*. As seen in Fig. **14**, the tires behave nonlinearly. The vehicle has a *4.45 cm* lateral movement during one sampling time of *1/24 s*. As the gap is *10 cm* for each side, this distance must be compensated by extra steering wheel adjustments. In Fig. **15**, the steering wheel plots are presented for a slower speed of *70 km/h* to show how accurate the driver model steering angle is when the lateral slip angle is small as shown in Fig. **16**. The difference is not considerable for slip angle of smaller than *1 deg*. where tires behave linearly.

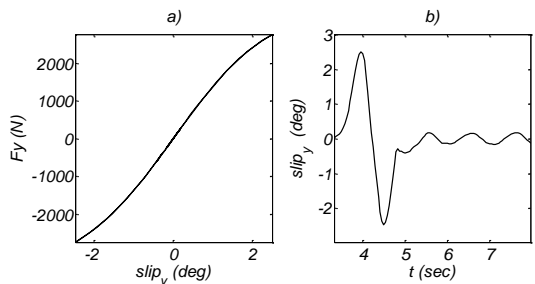

Fig. **14**. Tire lateral slip angles and forces in gap=20cm, d=50 (m), and V= 110 km/h. The tire behaves nonlinearly as seen in the first plot. a) lateral tire force-lateral slip. The tire behaves nonlinearly b) lateral slip-time

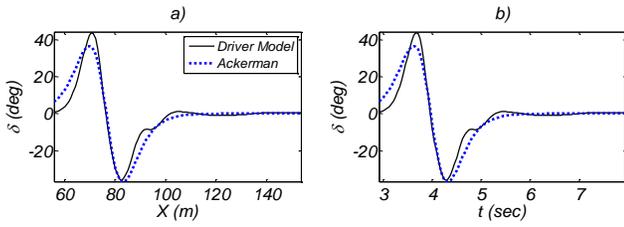

Fig. **15**. Steering wheel angle and steering wheel change in gap=20cm, d=50 (m), and V= 70 km/h. a) steering wheel angle-horizontal location b) steering wheel angle –time. The difference between Ackerman and driver model steering wheel angle is small.

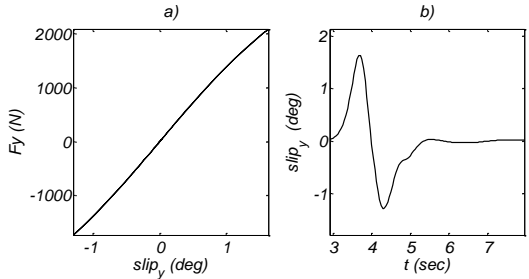

Fig. **16**. Steering wheel angle and steering wheel change in gap=20cm, d=50 (m), and V= 70 km/h. a) lateral tire force-lateral slip. The tire behaves linearly b) lateral slip-time

All in all, the driver model can show high performance even for tiny gaps, rapid lateral movements, and wide range of speeds.

The CPU time is an important factor for real time simulation of ADAS applications. The CPU time for each obstacle is less than *0.002 s* with an Intel i5 dual core 2.5 GHz CPU. This time is small enough for real-time applications if the sample time is more than *0.004 s*. In our simulation studies, the sample times shorter than 0.01*s* do not improve the simulation performance. Thus, the computational cost is not considerable.

*3.2 Positioning the vehicle at the presence of multiple obstacles*

This scenario includes obstacles with different sizes, longitudinal and lateral relative motions including upward, downward, forward, and backward relative motions.

In this test, the vehicle is surrounded by four different vehicles. Two vehicles travel at the right side of the vehicle. These two vehicles are a *5.5m×2.2m* van with and a *3.6m×1.6m* sedan. On the left side, there are a *8.6m×2.5m* bus and a *1.6m×0.4m* motorcycle. The speed of the vehicle is about *80 km/h*. The tracking path of the obstacles is formulated in Table 2.

Table 2
Paths of obstacles in second scenario

| | **Obstacle Path formula** | **relative motion** |
|---|---|---|
| **Bus** | $x(t) = 5 + x_V(t)$<br>$y(t) = 5$ | none |
| **Motorcycle** | $x(t) = 2 + x_V(t) + 2.3 \times sig(t, 4.34, 6.1)$<br>$y(t) = -5 + 6.6 \times sig(t, 4.34, 6.1)$ | Downward Backward |
| **Sedan** | $x(t) = -2 + x_V(t)$<br>$y(t) = -6 + 7 \times sig(t, 4.55, 6.1)$ | Upward forward |
| **Van** | $x(t) = -1 + x_V(t) - 2.25 \times sig(t, 4.55, 6.1)$<br>$y(t) = 6 - 2.6 \times sig(t, 4.55, 6.1)$ | Upward |

where $x_V$ is the horizontal location of the vehicle and the sigmoid function is defined as:

$$sig(t, a, b) = \frac{1}{1 + e^{-a(t-b)}} \qquad (47)$$

As shown in Fig. **17** the model can avoid colliding with all vehicles while four obstacles are getting closer to the vehicle as long as a tight space would exist. The steering wheel angle change for all obstacles is plotted in Fig. **18** . The semi-symmetric shape shows acceptable balance between left obstacles and right obstacles. The exerted steering wheel angle change and steering wheel angle are uniform and acceptable as shown in Fig. **19** and Fig. **20**.

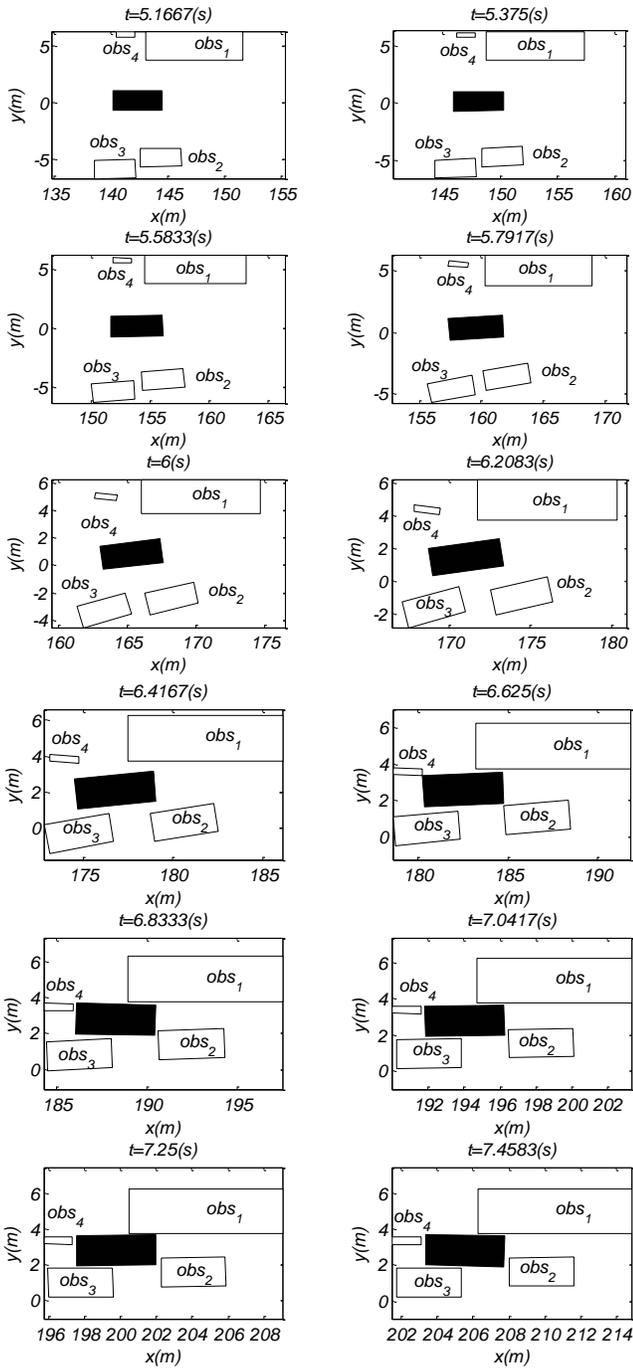

Fig. **17**. Traffic configuration including four obstacles of a bus, a van, a motorcycle, and a sedan at different times

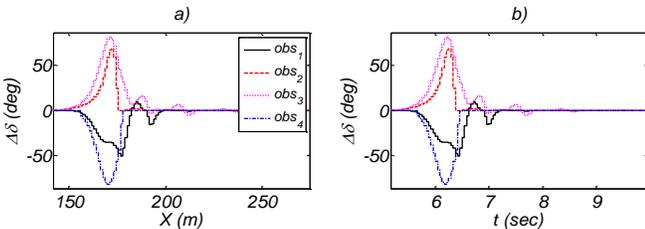

Fig. **18**. The steering wheel angle change for four obstacles in second scenario. a) Steering wheel angle change-horizontal location. b) Steering wheel angle change-time.

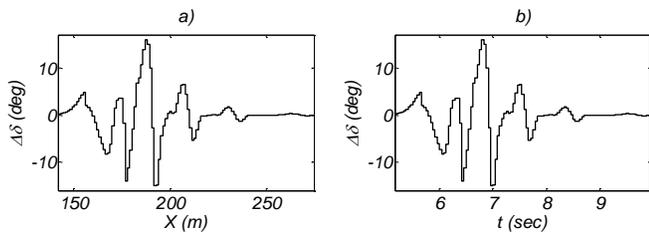

Fig. **19**. . The exerted steering wheel angle change in second scenario. a) Steering wheel angle change -horizontal location. b) Steering wheel angle change -time.

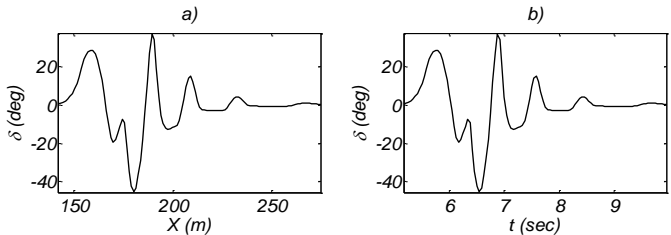

Fig. **20**. The steering wheel angle in the second scenario

## 4. Experimental test on a driving simulator

To fit driver model steering angle with human driver's steering angle, some cognitive and physical limitations should be modeled. Human parameters can model the most important physical and cognitive limitations cause human driver's inefficiency. In section 4.1, the human driver's inefficiency is modeled by parameterizing the driver model. The method of human parameter estimation is developed in section 4.2. In section 4.3, the experimental test is designed. Then, the results are analyzed for three drivers. It is demonstrated that the parameterized driver model can be matched with human drivers. Also, the human drivers' behavior can be analyzed based on the human parameters values.

### 4.1. Modeling human driver's inefficiency by parameterizing the driver model

The driver model steering angle is fitted to human drivers' steering angle using three human parameters. The human parameters include the maximum steering rate, the minimum noticeable difficulty threshold, and the sensitivity gain. If a human driver can turn steering wheel very fast, respond the small values of collision difficulty, and uses driver model steering gain $"K_s"$, it should have the efficiency of the driver model.

Based on physical and cognitive limitations, human drivers are not able to turn the steering wheel faster than their maximum steering rate $\dot{\delta}_{max}$. Also, they do not react to the task difficulty values smaller than the minimum noticeable difficulty threshold $TD_{min}$. Finally, the drivers' sensitivity toward task difficulty is not as same as the driver model. They may overestimate or underestimate the needed activation towards collision difficulty. This inefficiency is reflected by a sensitivity gain $K_{sen}$. Based on these factors, the parametric driver model can be formulated as:

$$\Delta \delta_m^* = K_{sen} K_s \left( TD - TD_{\min} \right) \qquad (48)$$

where $K_s$ is the steering control gain computed by equation (31), and $\Delta \delta_m^*$ is the uncapped driver model steering angle difference. The maximum steering rate that drivers can exert is denoted by $\dot{\delta}_{\max}$. Thus, the uncapped driver model steering angle difference should be bounded to estimate the driver model steering angle difference:

$$\Delta \delta_m = \begin{cases} \Delta \delta_m^* & \left| \Delta \delta_m^* \right| \le \dot{\delta}_{\max} T_s \\ \dot{\delta}_{\max} T_s sign(\Delta \delta_m^*) & \left| \Delta \delta_m^* \right| > \dot{\delta}_{\max} T_s \end{cases} \qquad (49)$$

where $T_s$ is the sampling period, and $\Delta \delta_m$ is the steering angle change of parameterized driver model. The following conditions should exist based on our expectation from a logical driver:

- The sensitivity gain should be nonzero and positive.
- The minimum noticeable difficulty threshold should be positive.

A negative sensitivity gain means the driver moves towards the most demanding obstacle instead of moving away. The logic of the driver model in this paper as described in section 2.4 requires the driver to avoid the most demanding obstacle. Several reasons may contribute to a driving behavior that results in approaching an obstacle. For example, if the driver cannot detect the most demanding obstacle, s/he may move towards it. Also, emotions like excessive fear and rage may cause incorrect decisions.

A negative minimum noticeable threshold means the driver moves away from an obstacle despite no collision difficulty. Such situations may happen if the driver makes decisions for overtaking and lane changing instead of avoiding collision.

Thus, the bounding conditions based on acceptable ranges of sensitivity gain and minimum noticeable difficulty threshold are as follows:

1. The driver should move away from the most demanding obstacle.
2. The driver should not move away from obstacles that have no task difficulty.

*4.2 Estimation of human parameters in the experimental tests*

The steering angle rate and difference, the position, the velocity, and the acceleration of the vehicle are used to estimate human driver parameters. The maximum steering rate can be found by defining the maximum absolute value of the driver's measured steering rate $\dot{\delta}_d$:

$$\dot{\delta}_{\max} = \max\left(\left|\dot{\delta}_d\right|\right) \tag{50}$$

Based on equations (48) and (49), the measured human driver's steering angle difference $^{\Delta\delta_d}$ smaller than the saturation limit is related to the sensitivity gain $K_{sen}$ and the minimum noticeable difficulty threshold $TD_{min}$. Accordingly, the measured steering angle difference greater than the saturation limit in equation (49) should not be used for parameter estimation. The extracted steering angle difference samples $^{\Delta\delta_d^*}$ are picked based on the following condition to estimate human driver parameters:

$$\Delta\delta_d^* = \left\{\Delta\delta_d \middle\| \left|\Delta\dot{\delta}_d\right| < T_s\dot{\delta}_{\max}\right\} \tag{51}$$

$\Delta\delta_m^*$ in equation (48) can be replaced by the extracted steering angle difference $\Delta\delta_d^*$ as follows:

$$\Delta\delta_d^* = K_s TD.K_{sen} - K_s.K_{sen}TD_{min} \tag{52}$$

The above equation can be rewritten in a linear form:

$$\begin{aligned}
\Delta\delta_d^* &= \phi\theta \\
\phi &= \begin{bmatrix} K_s TD & -K_s \end{bmatrix} \\
\theta &= \begin{bmatrix} K_{sen} \\ K_{sen}TD_{min} \end{bmatrix}
\end{aligned} \tag{53}$$

Thus, the parameters can be found using the pseudo inverse of matrix $\phi$:

$$\theta = \left(\phi^T\phi\right)^{-1}\phi^T\Delta\delta_d^* \tag{54}$$

Thus, the sensitivity gain $K_{sen}$ and minimum noticeable difficulty threshold $TD_{min}$ can be obtained by the two components of $\theta$. Note that if the sensitivity gain is zero, the minimum noticeable difficulty threshold cannot be computed. A zero sensitivity gain means the driver does not turn the steering wheel, thus the minimum difficulty at which the driver begins to react would be unmeasurable. This happens when the driver fails to detect an obstacle.

### 4.3. Experimental test and results

A simple test is designed on a sedan driving simulator to evaluate if the parameterized driver model is close to the human drivers 'behavior. The simulator is shown in Fig. **21** and the graphical environment of the test is shown in Fig. **22**. A "90-degree turn maneuver" is chosen for the test scenario as shown in Fig. **23**. The 8-meter-wide path includes a 200m straight road followed by a 90-degree turn with the inner and outer radii of *30m* and *38m*, respectively. The traveling speed is *60 km/h*. The drivers should avoid colliding with the road edge at both sides. The following considerations satisfy the bounded conditions of the parameterized driver model in the experiments:

1- Since steering is the dominant maneuver, the speed is maintained at 60 km/h during the tests and drivers are not allowed to use the accelerator or the brake pedals.

2- The estimated human parameters obtained from the tests are all non-negative as mentioned in section 4.1.

3- Detection errors are eliminated as obstacles are only road edges. Thus, all drivers can see the road edges several seconds before turning.

4- Drivers consider collision avoidance as the goal of driving during the tests. The mean radius of curvature is chosen as small as possible so that lane change will not be plausible for drivers.

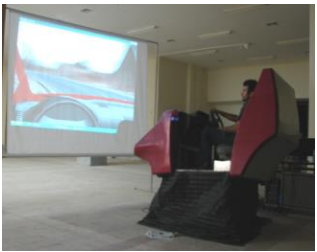

Fig. **21**. The driving simulator used for the tests

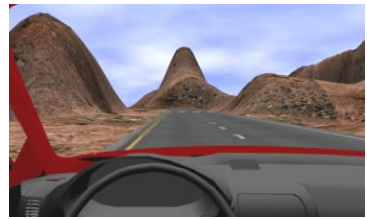

Fig. **22**. The graphical view of the exepriment

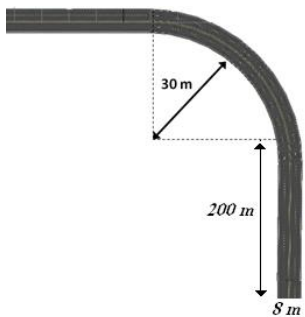

Fig. **23**. The maneuver path in driving simulator

Three drivers carried out the tests as case studies. The statistical data of steering wheel angle errors are shown in Table 3. As seen in Table 3, the mean of absolute error and the standard deviation of error are less than 9 degrees for all drivers. The normalized mean error is the ratio of the mean of absolute error to the maximum steering wheel angle during the 90-degree turn. The results for all three subjects are below 9%. These normalized errors are small enough indicating a good fit. The correlation coefficient is more than 0.98 that is close to 1. It means the shape of parametric driver model steering wheel angle and human drivers' are very similar. However the three samples are different in shape.

Table 3
The stochastic values of steering wheel angle errors

| Error statistical values: | Case subject 1 | Case subject 2 | Case subject 3 |
|---|---|---|---|
| Mean of absolute error (deg.) | 4.22 | 7.05 | 4.93 |
| STD of error (deg.) | 5.86 | 8.36 | 5.27 |
| Normalized mean of absolute error (×100) | 3.0% | 6.7% | 5.8% |
| STD of normalized error (×100) | 4.1% | 8.4% | 6.2% |
| Correlation coefficient | 0.996 | 0.988 | 0.997 |

The estimated parameters of drivers are given in Table 4. As seen in Table 4, drivers have different human factors. These parameters can justify their behaviors.

Table 4
The human driver parameters

| parameters: | Case subject 1 | Case subject 2 | Case subject 3 |
|---|---|---|---|
| $TD_{min}$ $(s^{-1})$ | 0.52 | 0.71 | 0.05 |
| $K_{sen}$ | 0.73 | 1.23 | 0.92 |
| Maximum steering rate $(deg./s)$ | 101 | 121 | 141 |

The steering angle of fitted driver model and human drivers are plotted in Fig. 24, Fig. 25, and Fig. 26.

Driver 1 does not turn the steering wheel when task difficulty is below $0.52s^{-1}$. The driver neglects the risk of potential collision when TTC is more than 2 s. Also, his sensitivity gain is small, *i.e. 0.73*. The driver should use large steering angle of more than 140 degrees to compensate for large minimum noticeable task-difficulty threshold and small sensitivity gain (Fig. 24).

Driver 2 exerts smaller steering wheel angles compared with driver 1 (Fig. 25). The minimum noticeable task-difficulty threshold is $0.71$ $s^{-1}$. As his maximum steering rate and sensitivity gains are more than those of driver 1, he uses lower steering angles. This happens despite the fact that the minimum noticeable task-difficulty threshold of driver 2 is slightly larger than driver 1. This means that the effects of the sensitivity gain and the maximum steering rate have exceeded the effect of the larger minimum noticeable task-difficulty threshold. The large sensitivity gains lead to overreactions. So, driver 2 uses large steering angles in the second overshoot, and a swinging behavior is observed.

Driver 3 uses the minimum steering angle range of less than 85 degrees as shown in Fig. 26. The turn radius is between *30 m* and *38 m*. The Ackerman steering wheel angles for these radii is *70 deg.* to *100 deg.* His minimum noticeable task-difficulty threshold is a small value of 0.05 $s^{-1}$. Thus, he needs only minor effort as he has reacted soon enough and has plenty of time to adjust to the ideal steering wheel angle trajectory.

His sensitivity gain is *0.92,* which is close to *1*. A sensitivity gain of 1 would indicate that the driver model and the human driver show the same sensitivity to task difficulty. For example, driver 3 behaves more similarly to an ideal driver model, which is used in the simulation studies of this paper. Also, his maximum steering angle rate is *141 deg./s* and he turns the wheel fast enough.

Driver 3 presents smaller steering wheel angle turn, less swinging behavior, and sooner reaction than drivers 1 and 2. It can be concluded that he is more skillful. Considering equation (41), when $K_s$=1, $TD_{min}$=0, and $\dot{\delta}_{max}$ is large enough, the human parameterized driver model and the ideal driver model used in simulation studies are the same. Driver 3 has the sensitivity gain of close to 1, the largest $\dot{\delta}_{max}$, and very small $TD_{min}$ of near zero. Thus, he behaves closely to our ideal driver model.

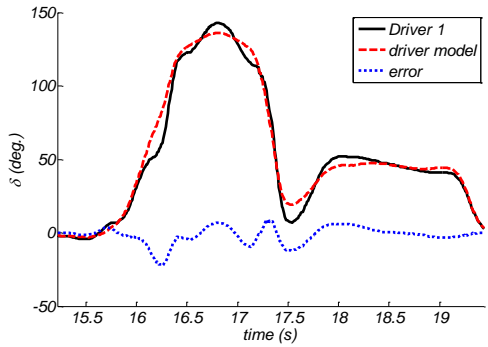

**Fig. 24**. Driver 1 and Model 1 steer angle

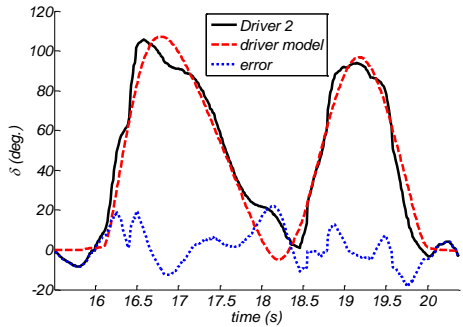

**Fig. 25**. Driver 2 and Model 2 steer angle

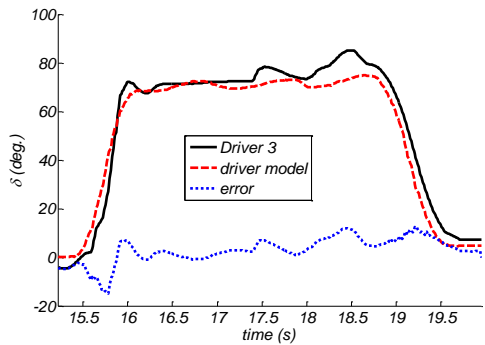

**Fig. 26**. Driver 3 and Model 3 steer angle

## 5. Conclusion

In this study, a psychologically plausible driver model for steering collision avoidance is presented. This model needs no gain tuning or algorithm shifting for different traffic configurations, obstacles, and speeds.

This model is based on a unified multidisciplinary approach. In this approach, a motivational model is mathematically formulated by control theory methods so that the resulted model is both stable and general. The task difficulty homeostasis theory as a motivational theory is formulated by Lyapunov stability method in control theory to create a simple, general, and stable model.

This model can avoid collision in small gaps and severe attacks (gap=20 cm, d=20 m). Also, the steering angle response is smooth and logical in comparison with the Ackerman estimation. In a general scenario, it shows satisfying results and avoids obstacles with different sizes and nonsynchronous movements. The CPU time for computation of each steering angle change is less than 0.002 seconds with an Intel i5 dual core 2.5 GHz CPU.

Using a driving simulator, the similarity of the model with human drivers is evaluated. By defining three human parameters, the model can fit driver steering responses with different shapes of steering responses. The mean of the error is less than 8% in steering angle. Also, the three human parameters used in this study, *i.e.* maximum steering rate, sensitivity gain, and minimum noticeable task difficulty threshold are directly related to human driver characteristics and can be used to judge driver performance.

All in all, a general, accurate, stable, and psychologically plausible model for steering collision avoidance is created so that the computational cost is low.

Some attractive application of this model will be developed. For example, the driving skills can be evaluated for both novice and professional racing drivers. Also a switching function can be designed using task difficulty to determine whether the driver is capable to control the vehicle in dangerous situation or not. This switching function may be used in ADAS to permit intelligent controller assist the incapable driver near accidents.

In our future research, we plan to develop a new method to include both longitudinal and lateral control to avoid collision. This model estimates task difficulty regarding various steering angles and speeds. Then, it makes the best decision so that task difficulty is acceptable and a higher speed is provided. This model would overcome more complicated scenarios and can be considered as a primitive general driver model.

### Acknowledgement

This paper is based upon work was supported by the Cognitive sciences and technologies Council of Iran (CSTC) under grant No. 1307.